\documentstyle[multicol,aps,epsfig]{revtex}

\begin{document}
\title{Enhanced quantized current driven by surface acoustic waves}
\author{J.~Ebbecke, G.~Bastian, M.~Bl\"ocker, K.~Pierz, and F.~J.~Ahlers}
\address{Physikalisch-Technische Bundesanstalt, Bundesallee 100, D-38116
Braunschweig, Germany}

\date{\today}
\maketitle
\begin{abstract}
We present the experimental realization of different approaches to
increase the amount of quantized current which is driven by
surface acoustic waves through split gate structures in a two
dimensional electron gas. Samples with driving frequencies of up
to 4.7~GHz have been fabricated without a deterioration of the
precision of the current steps, and a parallelization of two
channels with correspondingly doubled current values have been
achieved. We discuss theoretical and technological limitations of
these approaches for metrological applications as well as for
quantum logics.
\end{abstract}
\pacs{}

\begin{multicols}{2}

Within the last decade, circuits with single-electron-transistors
(SET) have been successfully fabricated for high frequencies up
to the 100~MHz range for electrometers \cite{Wahlgreen99}. In
application of SETs as electron-pumps e.g. for current standards,
however, the maximum operation frequency $f$ is always limited to
several 10~MHz. Only then the necessary tunneling events occur
with a sufficient probability and a high precision of current is
provided. The resulting small output currents in the order of
1~pA for a single device are both, difficult to measure with a
high absolute precision and ineligible to drive other devices
(e.g. quantum dot lasers \cite{Wiele98,Wiele99} or resistance
standards \cite{Klitzing80}). A single device also needs at least
2 gate contacts \cite{Lotkov00} and influences of local unstable
background changes need to be compensated \cite{Flensberg99b},
thus a parallelization of SET based devices to increase the
current output implies a delicate operation and to our knowledge
has not been realized yet.

In contrast to that, quantized acousto-electric currents driven by
surface acoustic waves (SAW) through a split-gate confined
potential within a 2DEG are by far larger due to the higher
possible operation frequencies of about 2.7~GHz realized up to
now \cite{Shilton96}. In this letter we show how such quantized
currents can be increased even further using higher operation
frequencies up to 4.7~GHz as well as by using parallelization of
split-gate structures.

Our samples were fabricated similarly to earlier works by Shilton
et al. \cite{Shilton96}. The inter-digital-transducers (IDTs) for
the highest frequencies were scaled down to a periodicity of
500~nm using a careful compensation of proximity effects for the
electron-beam exposure. The AlAs/GaAs heterostructures have
carrier mobilities of 300,000 to 650,000~cm$^2/$Vs and
concentrations of 2.6 to $4.2\times10^{11}~$cm$^{-2}$. This
results in mean free pathes above 3~$\mu$m that are much larger
than all fabricated channel lengths. These channels have been
defined by metallic Schottky-gates as well as by wet etched
trenches within the 2DEG \cite{Kristensen98b}.

\begin{figure}
\begin{center}
\epsfig{file=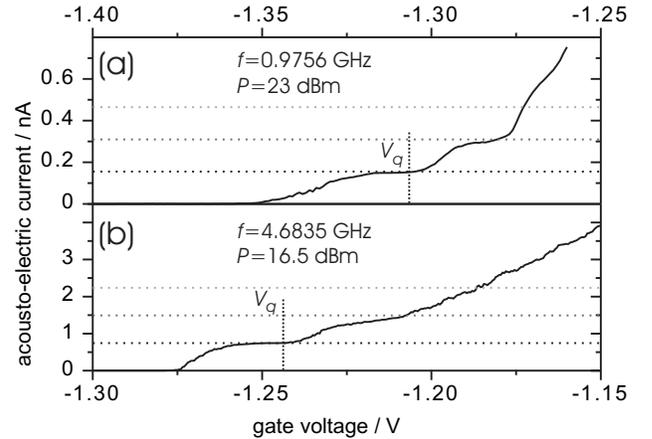}
\end{center}
\caption{Measured acousto-electric current versus voltage applied
to the rectangular shaped Schottky-split-gates at a temperature
of 1.4~K. The horizontal dashed lines indicate the values
$n\times e \times f$. a) Sample driven with $f$=0.9756~GHz and
split-gate dimensions of 800~nm$\times$ 2600~nm. b) At
$f$=4.6835~GHz a much higher rf-power $P$ is needed. The
dimensions of the split-gate are $700~nm\times 700~nm$.}
\label{fig1}
\end{figure}

There are several constraints for the operation frequency of SAW
current standards: the lower frequency limit is given by the
minimum SAW-induced confinement potential. Its increased width
leads to a lower energetical separation of the single electron
levels such that the quantization of the number of electrons per
cycle is washed out \cite{Comm04}. In addition to that half of the
SAW-wavelength must always be smaller than the channel length to
achieve quantized currents \cite{Ebbecke00}. In fig.\ref{fig1}a)
the acousto-electric current versus split-gate voltage $V_G$ is
shown for an SAW frequency of about 1~GHz. The first flat plateau
with quantized current $I=e\times f$ develops at split-gate
voltages around $V_q$ slightly below the pinch-off voltage. A
further decrease in operation frequency such that the SAW-driven
electrons can be detected with a comparable slow SET-electrometer
appears accomplishable. Samples with higher energetic level
spacing and also larger channel lengths are needed, whereas the
latter requires high mobility heterostructures to avoid
backscattering inside of the channels.

\begin{figure}
\begin{center}
\epsfig{file=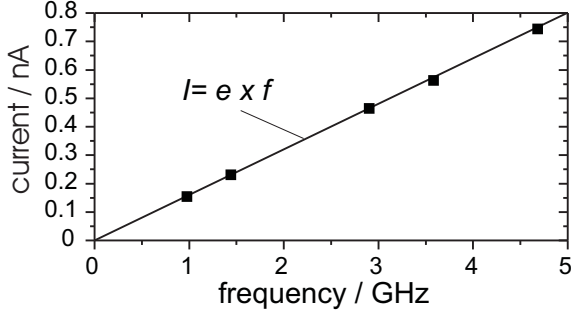}
\end{center}
\caption{The measured quantized acousto-electric current was
measured for 5 different frequencies. It depends linearly on the
driving frequency up to 4.7~GHz without an increase of deviations
from the quantized value.} \label{fig2}
\end{figure}

The high frequency limit is first of all set by technological
problems. The piezoelectric effect itself works up to the
Debye-frequency (far into the THz range), but the conversion of
rf-signals into a sufficiently large induced SAW potential within
a GaAs based 2DEG requires large area IDTs, which are difficult to
fabricate for high frequencies \cite{Willett95}. Our IDTs consist
of 70 finger pairs with a length of 80~$\mu$m and nominal widths
and separations down to 125~nm. The resulting current plateaus
for the corresponding frequency of 4.7~GHz can be seen in
fig.\ref{fig1}b). The flatness of the plateau
\begin{equation}\label{flatness}
 min\{dI_{SD}/dV_G\}=dI_{SD}/dV_G(V_G=V_q)\approx 0
\end{equation}
and the deviation of the absolute value at this gate-voltage from
the theoretical value
\begin{equation}\label{deviation}
 I(V_G=V_q)-n e f\approx0
\end{equation}
are not significantly worse compared to our devices with lower
operation frequencies. According to Flensberg et al.
\cite{Flensberg99} due to nonadiabaticity effects the precision
of the quantized currents should be reduced if
\begin{equation}\label{adiabaticity}
f\tau\ll 1
\end{equation}
with an estimated characteristic back-tunneling time
$\tau\approx$10~ps. Even at this high frequencies, where the
condition (\ref{adiabaticity}) is not strictly fulfilled
($f\tau=1/20$ for $f$=5~GHz) this kind of deviations appears to
be not dominant \cite{Comm01}. An overview over measured current
plateau values of samples with different operation frequencies is
depicted in fig.\ref{fig2}. Within the accuracy of our
measurements no significant deviation from the linear frequency
dependence of the quantized current values could be observed.

\begin{figure}
\begin{center}
\epsfig{file=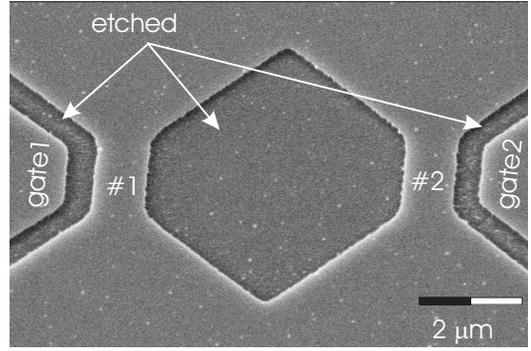}
\end{center}
\caption{The double SAW-split-gate device is patterned with
etched trenches of 600~nm width and 90~nm depth. The geometrical
length and width of each of the channels ($\sharp1$, $\sharp2$) is
2~$\mu$m and 960~nm, respectively. The width of the channel can
be changed with voltage applied to the side-gates.} \label{fig3}
\end{figure}

In addition to samples with enhanced IDTs we also have fabricated
samples in which two split-gate structures work in parallel. The
inner geometry of such samples can be seen in the SEM-picture in
fig.\ref{fig3}. A shallow etching technique has been used to
remove the doped GaAs layer locally in order to deplete the 2DEG
and thus isolate the channels from the two side-gate areas
\cite{Pepper00}. The distance $X=5~\mu$m between the two channels
has been chosen large enough such that the Coulomb energy between
two passing electrons

\begin{equation}\label{Coulomb-repulsion}
E_C=\int_X^{\infty}
\frac{1}{4\pi\epsilon_0\epsilon_r}\frac{e^2}{r^2}dr=0.02~\textrm{meV}
\end{equation}

is negligible compared with other energy scales in the device.
Therefore Coulomb-drag effects \cite{Rojo99} were not utilized to
modify the device's properties. Also interference effects are
assumed to play a crucial role only, if the length of a loop
through the two channels is significantly shorter.

\begin{figure}
\begin{center}
\epsfig{file=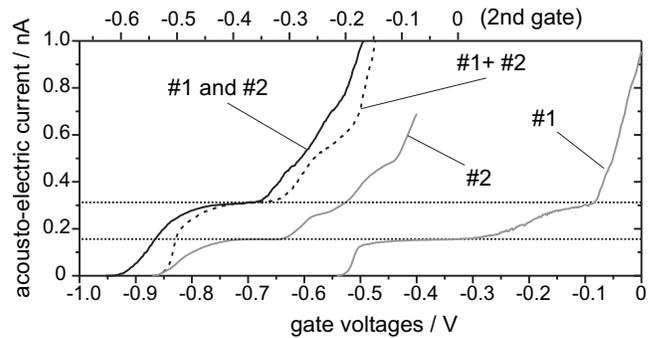}
\end{center}
\caption{Acousto-electric current measured for each of the
channels ($\sharp1$, $\sharp2$) versus gate voltage applied to
the respective nearby gate (lower voltage scale). The calculated
sum of currents through the two single channels (dashed line) and
the measured parallel current ($\sharp$1 and $\sharp$2) exhibits a
first current plateau at 2$\times e \times f$, whereas the voltage
applied to the second side-gate is shown in the upper scale. The
applied rf-frequency and -power is $\approx$0.976~GHz and 13~dBm,
respectively.} \label{fig4}
\end{figure}

First, the measurement parameters for each channel $i$ were
independently optimized, namely $V_{q,i}$, the frequency and
rf-power, while the other channel was completely pinched off. In
a second step both channels were opened slightly by applying
voltages $V_{q,1}$ and $V_{q,2}$ to allow parallel transport of
electrons within each cycle, while the influence of the
respective more distant side-gates was also taken into account.
In fig.\ref{fig4} the measured quantized current for the two
individual channels, their algebraic sum and the measured
parallel current is depicted for the above sample. The precision
of the doubled quantized current is not reduced compared to the
single channel value, which indicates the applicability and
benefit of this parallelization for metrology \cite{Comm03}.

Recently, a parallel connection of SAW-driven electrons has also
been proposed for quantum computation \cite{Barnes00}. One major
benefit of this concept over other proposals lies in the high
repetition rate of individual qubit operations leading to a
reduction of errors. As shown in fig.\ref{fig4}, the first step
towards a realization of this concept, namely the parallelization
of SAW-driven electrons, is possible if the gate-voltages are
chosen properly. Higher operation frequencies, as shown in the
previous section, might be advantageous for this application as
well.

Other approaches to increase the amount of quantized current were
not successful yet: the use of a different piezoelectric
substrate than GaAs with a higher sound velocity $v_S$ could
easily provide a higher frequency ($f=v_S/\lambda$) and a higher
amplitude for a given periodicity $\lambda$ of the IDTs. Also the
application of a higher harmonic ($n\times f$) to split-finger
IDTs seems promising, but needs high power levels that are not
easy to provide. Until now, also pumping of more than one
electron per cycle ($n\times e$) \cite{Gumbs99} could not be
realized with sufficient accuracy. We also could not confirm
\cite{Cunningham99} the theoretical assumption, that the
energetic level spacing (caused by Coulomb blockade) is constant
and therefore the width of all current plateaus, too \cite
{Flensberg99}. In contrast, we sometimes observe only one or two
plateaus.

In conclusion, different approaches for the enhancement of the
quantized current value in SAW driven structures have been
experimentally studied and realized. It was shown that using
higher frequencies of up to 4.7~GHz as well as a parallel
connection of split-gates the device could be significantly
improved. Using the presented techniques, other related effects
like SAW-quantum computing, SAW-pumped lasers and
SAW-current-driven resistance standards get closer to realization.

\acknowledgments We would like to thank T.~Weimann for fruitful
discussions, H.~Marx for providing the 2DEGs and the Deutsche
Forschungsgemeinschaft (DFG Ah79/2-1) for financial support.

\end{multicols}

\begin{references}
\bibitem{Comm04}Similar constraints are given if the Coulomb blockade
energy and not the level quantization determines the quantization
of the number of electrons.
\end{references}
\end{document}